\expandafter \def \csname CHAPLABELintro\endcsname {1}
\expandafter \def \csname EQLABELePerDef\endcsname {1.1?}
\expandafter \def \csname EQLABELeOM\endcsname {1.2?}
\expandafter \def \csname EQLABELeLaplace\endcsname {1.3?}
\expandafter \def \csname EQLABELprotosp\endcsname {1.4?}
\expandafter \def \csname CHAPLABELghs\endcsname {2}
\expandafter \def \csname EQLABELnu2\endcsname {2.1?}
\expandafter \def \csname EQLABELlindep\endcsname {2.2?}
\expandafter \def \csname EQLABELsys\endcsname {2.3?}
\expandafter \def \csname EQLABELzetadef\endcsname {2.4?}
\expandafter \def \csname CHAPLABELsex\endcsname {3}
\expandafter \def \csname EQLABELDtorus\endcsname {3.1?}
\expandafter \def \csname EQLABELZitorus\endcsname {3.2?}
\expandafter \def \csname EQLABELDtorus\endcsname {3.3?}
\expandafter \def \csname EQLABELPistorus\endcsname {3.4?}
\expandafter \def \csname EQLABELperdef\endcsname {3.5?}
\expandafter \def \csname EQLABELperdefII\endcsname {3.6?}
\expandafter \def \csname EQLABELsymtorI\endcsname {3.7?}
\expandafter \def \csname EQLABELZsymtorI\endcsname {3.8?}
\expandafter \def \csname EQLABELsymtorII\endcsname {3.9?}
\expandafter \def \csname EQLABELZsymtorII\endcsname {3.10?}
\expandafter \def \csname EQLABELDsymtor\endcsname {3.11?}
\expandafter \def \csname EQLABELFZ\endcsname {3.12?}
\expandafter \def \csname EQLABELFZcalc\endcsname {3.13?}
\expandafter \def \csname CHAPLABELexample\endcsname {4}
\expandafter \def \csname EQLABELF\endcsname {4.1?}
\expandafter \def \csname EQLABELFII\endcsname {4.2?}
\expandafter \def \csname EQLABELpfexample\endcsname {4.3?}
\expandafter \def \csname CHAPLABELexampleIII\endcsname {5}
\expandafter \def \csname EQLABELd3\endcsname {5.1?}
\expandafter \def \csname EQLABELindeqsol\endcsname {5.2?}
\expandafter \def \csname EQLABELindeq\endcsname {5.3?}
\expandafter \def \csname EQLABELzeta2\endcsname {5.4?}
\expandafter \def \csname EQLABELd2\endcsname {5.5?}
\expandafter \def \csname EQLABELspt2\endcsname {5.6?}
\expandafter \def \csname EQLABELspk3\endcsname {5.7?}
\expandafter \def \csname EQLABELspcy\endcsname {5.8?}
\expandafter \def \csname EQLABELBarnes\endcsname {5.9?}
\expandafter \def \csname EQLABELksum\endcsname {5.10?}

\font\eightrm=cmr8 at 8pt

\font\seventeenrm=cmr17 at 17pt
\font\twentyonerm=cmr17 at 21pt

\font\ss=cmss10

\font\csc=cmcsc10

\font\twelvecal=cmsy10 at 12pt

\font\twelvemath=cmmi12

\font\seventeenbold=cmbx7 at 17pt

\font\fively=lasy5
\font\sevenly=lasy7
\font\tenly=lasy10

\textfont10=\tenly
\scriptfont10=\sevenly
\scriptscriptfont10=\fively
\magnification=1200
\parskip=10pt
\parindent=20pt
\def\today{\ifcase\month\or January\or February\or March\or April\or May\or
June
       \or July\or August\or September\or October\or November\or December\fi
       \space\number\day, \number\year}

\def\title#1{\footline={\ifnum\pageno<2\hfil
       \else\hss\tenrm\folio\hss\fi}\vskip1truein\centerline{{#1}
       \footnote{\raise1ex\hbox{*}}{\eightrm Supported in part
       by the Robert A. Welch Foundation and N.S.F. Grants
       PHY-880637 and\break PHY-8605978.}}}

\def\Z{\hfill\break}

\def\abstract#1{\centerline{\bf ABSTRACT}\vskip.2truein{\narrower\noindent#1
       \smallskip}}

\def\runninghead#1#2{\voffset=2\baselineskip\nopagenumbers
       \headline={\ifodd\pageno\rightheadline\else \leftheadline\fi}
       \def\rightheadline{{\sl#1}\hfill{\rm\folio}}
       \def\leftheadline{{\rm\folio}\hfill{\sl#2}}}

\newcount\footnoteno
\def\Footnote#1{\advance\footnoteno by 1
                \let\SF=\empty
                \ifhmode\edef\SF{\spacefactor=\the\spacefactor}\/\fi
                $^{\the\footnoteno}$\ignorespaces
                \SF\vfootnote{$^{\the\footnoteno}$}{#1}}

\def\place#1#2#3{\vbox to0pt{\kern-\parskip\kern-7pt
                             \kern-#2truein\hbox{\kern#1truein #3}
                             \vss}\nointerlineskip}
\def\figurecaption#1#2{\kern.75truein\vbox{\hsize=5truein\noindent{\bf Figure
    \figlabel{#1}:} #2}}
\def\tablecaption#1#2{\kern.75truein\lower12truept\hbox{\vbox{\hsize=5truein
    \noindent{\bf Table\hskip5truept\tablabel{#1}:} #2}}}
\def\boxed#1{\lower3pt\hbox{
                       \vbox{\hrule\hbox{\vrule

\vbox{\kern2pt\hbox{\kern3pt#1\kern3pt}\kern3pt}\vrule}
                         \hrule}}}
\def\a{\alpha}

\def\g{\gamma}\def\G{\Gamma}
\def\d{\delta}\def\D{\Delta}
\def\e{\epsilon}
\def\z{\zeta}

\def\l{\lambda}

\def\vp{\varpi}

\def\ph{\phi}

\def\ps{\psi}
\def\o{\omega}\def\O{\Omega}

\def\ca#1{\relax\ifmmode {{\cal #1}}\else $\cal #1$\fi}

\def\calb{{\cal B}}

\def\calm{{\cal M}}

\def\inbar{\vrule height1.5ex width.4pt depth0pt}
\def\IB{\relax{\rm I\kern-.18em B}}
\def\IC{\relax\hbox{\kern.25em$\inbar\kern-.3em{\rm C}$}}
\def\ID{\relax{\rm I\kern-.18em D}}
\def\IE{\relax{\rm I\kern-.18em E}}
\def\IF{\relax{\rm I\kern-.18em F}}
\def\IG{\relax\hbox{\kern.25em$\inbar\kern-.3em{\rm G}$}}
\def\IH{\relax{\rm I\kern-.18em H}}
\def\II{\relax{\rm I\kern-.18em I}}
\def\IK{\relax{\rm I\kern-.18em K}}
\def\IL{\relax{\rm I\kern-.18em L}}
\def\IM{\relax{\rm I\kern-.18em M}}
\def\IN{\relax{\rm I\kern-.18em N}}
\def\IO{\relax\hbox{\kern.25em$\inbar\kern-.3em{\rm O}$}}
\def\IP{\relax{\rm I\kern-.18em P}}
\def\IQ{\relax\hbox{\kern.25em$\inbar\kern-.3em{\rm Q}$}}
\def\IR{\relax{\rm I\kern-.18em R}}
\def\IZ{\relax\ifmmode\hbox{\ss Z\kern-.4em Z}\else{\ss Z\kern-.4em Z}\fi}
\def\IGa{\relax{\rm I}\kern-.18em\Gamma}
\def\IPi{\relax{\rm I}\kern-.18em\Pi}
\def\ITh{\relax\hbox{\kern.25em$\inbar\kern-.3em\Theta$}}
\def\IOm{\relax\thinspace\inbar\kern1.95pt\inbar\kern-5.525pt\Omega}


\def\ie{{\it i.e.,\ \/}}

\def\noblackboxes{\overfullrule=0pt}
\def\define{\buildrel\rm def\over =}

\def\cym{Calabi--Yau manifold}
\def\cys{Calabi--Yau manifolds}

\def\K{K\"ahler}

\def\H#1#2{\relax\ifmmode {H^{#1#2}}\else $H^{#1 #2}$\fi}
\def\M{\relax\ifmmode{\calm}\else $\calm$\fi}

\def\Bigcheck{\lower3.8pt\hbox{\smash{\hbox{{\twentyonerm \v{}}}}}}
\def\bigboldcheck{\smash{\hbox{{\seventeenbold\v{}}}}}

\def\Bighat{\lower3.8pt\hbox{\smash{\hbox{{\twentyonerm \^{}}}}}}

\def\Msharp{\relax\ifmmode{\calm^\sharp}\else $\smash{\calm^\sharp}$\fi}
\def\Mflat{\relax\ifmmode{\calm^\flat}\else $\smash{\calm^\flat}$\fi}
\def\preMcheck{\kern2pt\hbox{\Bigcheck\kern-12pt{$\cal M$}}}
\def\Mcheck{\relax\ifmmode\preMcheck\else $\preMcheck$\fi}
\def\preMhat{\kern2pt\hbox{\Bighat\kern-12pt{$\cal M$}}}
\def\Mhat{\relax\ifmmode\preMhat\else $\preMhat$\fi}

\def\Bsharp{\relax\ifmmode{\calb^\sharp}\else $\calb^\sharp$\fi}
\def\Bflat{\relax\ifmmode{\calb^\flat}\else $\calb^\flat$ \fi}
\def\preBcheck{\hbox{\Bigcheck\kern-9pt{$\cal B$}}}
\def\Bcheck{\relax\ifmmode\preBcheck\else $\preBcheck$\fi}
\def\preBhat{\hbox{\Bighat\kern-9pt{$\cal B$}}}
\def\Bhat{\relax\ifmmode\preBhat\else $\preBhat$\fi}

\def\figBcheck{\kern3pt\hbox{\raise1pt\hbox{\bigboldcheck}\kern-11pt
    {\twelvecal B}}}
\def\figBsharp{{\twelvecal B}\raise5pt\hbox{$\twelvemath\sharp$}}
\def\figBflat{{\twelvecal B}\raise5pt\hbox{$\twelvemath\flat$}}

\def\gcheck{\hbox{\lower2.5pt\hbox{\Bigcheck}\kern-8pt$\g$}}
\def\lhat{\hbox{\raise.5pt\hbox{\Bighat}\kern-8pt$\l$}}

\def\Fcheck{\kern2pt\hbox{\raise1pt\hbox{\Bigcheck}\kern-10pt{$\cal F$}}}
\def\Fhat{\kern2pt\hbox{\raise1pt\hbox{\Bighat}\kern-10pt{$\cal F$}}}

\def\cp#1{\relax\ifmmode {\IP\kern-2pt{}_{#1}}\else $\IP\kern-2pt{}_{#1}$\fi}
\def\h#1#2{\relax\ifmmode {b_{#1#2}}\else $b_{#1#2}$\fi}
\def\Z{\hfill\break}

\def\frac#1#2{{#1\over #2}}

\def\pd#1#2{{\partial #1\over\partial #2}}

\def\cone{\relax\thinspace\hbox{$<\kern-.8em{)}$}}
\mathchardef\mho"0A30

\def\Psymbol#1#2{\hbox{\twelvecal\char'120}\left\{
                 \matrix{0&\infty&1\cr #1}\hskip8truept\matrix{#2}\right\}}

\def\-{\hphantom{-}}


\def\npb#1{Nucl.\ Phys.\ {\bf B#1}}
\def\prl#1{Phys. Rev. Lett. {\bf #1}}
\def\cmp#1{Commun. Math. Phys. {\bf #1}}
\def\plb#1{Phys. Lett. {\bf #1B}}


\def\picture #1 by #2 (#3){\vbox to #2{\hrule width #1 height 0pt depth 0pt
                                       \vfill\special{picture #3}}}
\def\scaledpicture #1 by #2 (#3 scaled #4){{\dimen0=#1 \dimen1=#2
           \divide\dimen0 by 1000 \multiply\dimen0 by #4
            \divide\dimen1 by 1000 \multiply\dimen1 by #4
            \picture \dimen0 by \dimen1 (#3 scaled #4)}}
\def\illustration #1 by #2 (#3){\vbox to #2{\hrule width #1 height 0pt depth
0pt
                                       \vfill\special{illustration #3}}}
\def\scaledillustration #1 by #2 (#3 scaled #4){{\dimen0=#1 \dimen1=#2
           \divide\dimen0 by 1000 \multiply\dimen0 by #4
            \divide\dimen1 by 1000 \multiply\dimen1 by #4
            \illustration \dimen0 by \dimen1 (#3 scaled #4)}}


\def\delaOssa{\nobreak\vskip1truein\hbox to\hsize
       {\hskip 4truein Xenia de la Ossa\hfill}}

\def\hoy{\number\day\space de \ifcase\month\or enero\or febrero\or marzo\or
       abril\or mayo\or junio\or julio\or agosto\or septiembre\or octubre\or
       noviembre\or diciembre\fi\space de \number\year}


\newif\ifproofmode
\proofmodefalse

\newif\ifforwardreference
\forwardreferencefalse

\newif\ifchapternumbers
\chapternumbersfalse

\newif\ifcontinuousnumbering
\continuousnumberingfalse

\newif\iffigurechapternumbers
\figurechapternumbersfalse

\newif\ifcontinuousfigurenumbering
\continuousfigurenumberingfalse

\newif\iftablechapternumbers
\tablechapternumbersfalse

\newif\ifcontinuoustablenumbering
\continuoustablenumberingfalse

\font\eqsixrm=cmr6

\def\marginstyle{\eqsixrm}

\newtoks\chapletter
\newcount\chapno
\newcount\eqlabelno
\newcount\figureno
\newcount\tableno

\chapno=0
\eqlabelno=0
\figureno=0
\tableno=0

\def\chapfolio{\ifnum\chapno>0 \the\chapno\else\the\chapletter\fi}

\def\bumpchapno{\ifnum\chapno>-1 \global\advance\chapno by 1
\else\global\advance\chapno by -1 \setletter\chapno\fi
\ifcontinuousnumbering\else\global\eqlabelno=0 \fi
\ifcontinuousfigurenumbering\else\global\figureno=0 \fi
\ifcontinuoustablenumbering\else\global\tableno=0 \fi}

\def\setletter#1{\ifcase-#1{}\or{}%
\or\global\chapletter={A}%
\or\global\chapletter={B}%
\or\global\chapletter={C}%
\or\global\chapletter={D}%
\or\global\chapletter={E}%
\or\global\chapletter={F}%
\or\global\chapletter={G}%
\or\global\chapletter={H}%
\or\global\chapletter={I}%
\or\global\chapletter={J}%
\or\global\chapletter={K}%
\or\global\chapletter={L}%
\or\global\chapletter={M}%
\or\global\chapletter={N}%
\or\global\chapletter={O}%
\or\global\chapletter={P}%
\or\global\chapletter={Q}%
\or\global\chapletter={R}%
\or\global\chapletter={S}%
\or\global\chapletter={T}%
\or\global\chapletter={U}%
\or\global\chapletter={V}%
\or\global\chapletter={W}%
\or\global\chapletter={X}%
\or\global\chapletter={Y}%
\or\global\chapletter={Z}\fi}

\def\tempsetletter#1{\ifcase-#1{}\or{}%
\or\global\chapletter={A}%
\or\global\chapletter={B}%
\or\global\chapletter={C}%
\or\global\chapletter={D}%
\or\global\chapletter={E}%
\or\global\chapletter={F}%
\or\global\chapletter={G}%
\or\global\chapletter={H}%
\or\global\chapletter={I}%
\or\global\chapletter={J}%
\or\global\chapletter={K}%
\or\global\chapletter={L}%
\or\global\chapletter={M}%
\or\global\chapletter={N}%
\or\global\chapletter={O}%
\or\global\chapletter={P}%
\or\global\chapletter={Q}%
\or\global\chapletter={R}%
\or\global\chapletter={S}%
\or\global\chapletter={T}%
\or\global\chapletter={U}%
\or\global\chapletter={V}%
\or\global\chapletter={W}%
\or\global\chapletter={X}%
\or\global\chapletter={Y}%
\or\global\chapletter={Z}\fi}

\def\chapshow#1{\ifnum#1>0 \relax#1%
\else{\tempsetletter{\number#1}\chapno=#1\chapfolio}\fi}

\def\chaplabel#1{\bumpchapno\ifproofmode\ifforwardreference
\immediate\write1{\noexpand\expandafter\noexpand\def
\noexpand\csname CHAPLABEL#1\endcsname{\the\chapno}}\fi\fi
\global\expandafter\edef\csname CHAPLABEL#1\endcsname
{\the\chapno}\ifproofmode\llap{\hbox{\marginstyle #1\ }}\fi\chapfolio}

\def\chapref#1{\ifundefined{CHAPLABEL#1}??\ifproofmode\ifforwardreference%
\else\write16{ ***Undefined Chapter Reference #1*** }\fi
\else\write16{ ***Undefined Chapter Reference #1*** }\fi
\else\edef\LABxx{\getlabel{CHAPLABEL#1}}\chapshow\LABxx\fi
\ifproofmode\write2{Chapter #1}\fi}

\def\eqnum{\global\advance\eqlabelno by 1
\eqno(\ifchapternumbers\chapfolio.\fi\the\eqlabelno)}

\def\eqlabel#1{\global\advance\eqlabelno by 1 \ifproofmode\ifforwardreference
\immediate\write1{\noexpand\expandafter\noexpand\def
\noexpand\csname EQLABEL#1\endcsname{\the\chapno.\the\eqlabelno?}}\fi\fi
\global\expandafter\edef\csname EQLABEL#1\endcsname
{\the\chapno.\the\eqlabelno?}\eqno(\ifchapternumbers\chapfolio.\fi
\the\eqlabelno)\ifproofmode\rlap{\hbox{\marginstyle #1}}\fi}

\def\eqalignnum{\global\advance\eqlabelno by 1
&(\ifchapternumbers\chapfolio.\fi\the\eqlabelno)}

\def\eqalignlabel#1{\global\advance\eqlabelno by 1 \ifproofmode
\ifforwardreference\immediate\write1{\noexpand\expandafter\noexpand\def
\noexpand\csname EQLABEL#1\endcsname{\the\chapno.\the\eqlabelno?}}\fi\fi
\global\expandafter\edef\csname EQLABEL#1\endcsname
{\the\chapno.\the\eqlabelno?}&(\ifchapternumbers\chapfolio.\fi
\the\eqlabelno)\ifproofmode\rlap{\hbox{\marginstyle #1}}\fi}

\def\eqref#1{\hbox{(\ifundefined{EQLABEL#1}***)\ifproofmode\ifforwardreference%
\else\write16{ ***Undefined Equation Reference #1*** }\fi
\else\write16{ ***Undefined Equation Reference #1*** }\fi
\else\edef\LABxx{\getlabel{EQLABEL#1}}%
\def\LAByy{\expandafter\stripchap\LABxx}\ifchapternumbers%
\chapshow{\LAByy}.\expandafter\stripeq\LABxx%
\else\ifnum\number\LAByy=\chapno\relax\expandafter\stripeq\LABxx%
\else\chapshow{\LAByy}.\expandafter\stripeq\LABxx\fi\fi)\fi}%
\ifproofmode\write2{Equation #1}\fi}

\def\fignum{\global\advance\figureno by 1
\relax\iffigurechapternumbers\chapfolio.\fi\the\figureno}

\def\figlabel#1{\global\advance\figureno by 1
\relax\ifproofmode\ifforwardreference
\immediate\write1{\noexpand\expandafter\noexpand\def
\noexpand\csname FIGLABEL#1\endcsname{\the\chapno.\the\figureno?}}\fi\fi
\global\expandafter\edef\csname FIGLABEL#1\endcsname
{\the\chapno.\the\figureno?}\iffigurechapternumbers\chapfolio.\fi
\ifproofmode\llap{\hbox{\marginstyle#1
\kern1.2truein}}\relax\fi\the\figureno}

\def\figref#1{\hbox{%
\ifundefined{FIGLABEL#1}!!!!\ifproofmode\ifforwardreference%
\else\write16{ ***Undefined Figure Reference #1*** }\fi
\else\write16{ ***Undefined Figure Reference #1*** }\fi
\else\edef\LABxx{\getlabel{FIGLABEL#1}}%
\def\LAByy{\expandafter\stripchap\LABxx}\iffigurechapternumbers%
\chapshow{\LAByy}.\expandafter\stripeq\LABxx%
\else\ifnum \number\LAByy=\chapno\relax\expandafter\stripeq\LABxx%
\else\chapshow{\LAByy}.\expandafter\stripeq\LABxx\fi\fi\fi}%
\ifproofmode\write2{Figure #1}\fi}

\def\tabnum{\global\advance\tableno by 1
\relax\iftablechapternumbers\chapfolio.\fi\the\tableno}

\def\tablabel#1{\global\advance\tableno by 1
\relax\ifproofmode\ifforwardreference
\immediate\write1{\noexpand\expandafter\noexpand\def
\noexpand\csname TABLABEL#1\endcsname{\the\chapno.\the\tableno?}}\fi\fi
\global\expandafter\edef\csname TABLABEL#1\endcsname
{\the\chapno.\the\tableno?}\iftablechapternumbers\chapfolio.\fi
\ifproofmode\llap{\hbox{\marginstyle#1
\kern1.2truein}}\relax\fi\the\tableno}

\def\tabref#1{\hbox{%
\ifundefined{TABLABEL#1}!!!!\ifproofmode\ifforwardreference%
\else\write16{ ***Undefined Table Reference #1*** }\fi
\else\write16{ ***Undefined Table Reference #1*** }\fi
\else\edef\LABtt{\getlabel{TABLABEL#1}}%
\def\LABTT{\expandafter\stripchap\LABtt}\iftablechapternumbers%
\chapshow{\LABTT}.\expandafter\stripeq\LABtt%
\else\ifnum\number\LABTT=\chapno\relax\expandafter\stripeq\LABtt%
\else\chapshow{\LABTT}.\expandafter\stripeq\LABtt\fi\fi\fi}%
\ifproofmode\write2{Table#1}\fi}

\newdimen\sectionskip     \sectionskip=20truept
\newcount\sectno
\def\section#1#2{\sectno=0 \null\vskip\sectionskip
    \centerline{\chaplabel{#1}.~~{\bf#2}}\nobreak\vskip.2truein
    \noindent\ignorespaces}

\def\advancesectno{\global\advance\sectno by 1}
\def\sectfolio{\number\sectno}
\def\subsection#1{\goodbreak\advancesectno\null\vskip10pt
                  \noindent\chapfolio.~\sectfolio.~{\bf #1}
                  \nobreak\vskip.05truein\noindent\ignorespaces}

\def\uttg#1{\null\vskip.1truein
    \ifproofmode \line{\hfill{\bf Draft}:
    UTTG--{#1}--\number\year}\line{\hfill\today}
    \else \line{\hfill UTTG--{#1}--\number\year}
    \line{\hfill\ifcase\month\or January\or February\or March\or April\or
May\or June
    \or July\or August\or September\or October\or November\or December\fi
    \space\number\year}\fi}

\def\getlabel#1{\csname#1\endcsname}
\def\ifundefined#1{\expandafter\ifx\csname#1\endcsname\relax}
\def\stripchap#1.#2?{#1}
\def\stripeq#1.#2?{#2}

%
\catcode`@=11 
\def\space@ver#1{\let\@sf=\empty\ifmmode#1\else\ifhmode%
\edef\@sf{\spacefactor=\the\spacefactor}\unskip${}#1$\relax\fi\fi}
\newcount\referencecount     \referencecount=0
\newif\ifreferenceopen       \newwrite\referencewrite
\newtoks\rw@toks
\def\refmark#1{\relax[#1]}
\def\refend{\refmark{\number\referencecount}}
\newcount\lastrefsbegincount \lastrefsbegincount=0
\def\refsend{\refmark{\count255=\referencecount%
\advance\count255 by -\lastrefsbegincount%
\ifcase\count255 \number\referencecount%
\or\number\lastrefsbegincount,\number\referencecount%
\else\number\lastrefsbegincount-\number\referencecount\fi}}
\def\refch@ck{\chardef\rw@write=\referencewrite
\ifreferenceopen\else\referenceopentrue
\immediate\openout\referencewrite=referenc.texauxil \fi}
%
{\catcode`\^^M=\active 
  \gdef\obeyendofline{\catcode`\^^M\active \let^^M\ }}%
%
{\catcode`\^^M=\active 
  \gdef\ignoreendofline{\catcode`\^^M=5}}
{\obeyendofline\gdef\rw@start#1{\def\t@st{#1}\ifx\t@st\blankend%
\endgroup\@sf\relax\else\ifx\t@st\bl@nkend\endgroup\@sf\relax%
\else\rw@begin#1
\backtotext
\fi\fi}}
{\obeyendofline\gdef\rw@begin#1
{\def\n@xt{#1}\rw@toks={#1}\relax%
\rw@next}}
\def\blankend{}
{\obeylines\gdef\bl@nkend{
}}
\newif\iffirstrefline  \firstreflinetrue
\def\rwr@teswitch{\ifx\n@xt\blankend\let\n@xt=\rw@begin%
\else\iffirstrefline\global\firstreflinefalse%
\immediate\write\rw@write{\noexpand\obeyendofline\the\rw@toks}%
\let\n@xt=\rw@begin%
\else\ifx\n@xt\rw@@d \def\n@xt{\immediate\write\rw@write{%
\noexpand\ignoreendofline}\endgroup\@sf}%
\else\immediate\write\rw@write{\the\rw@toks}%
\let\n@xt=\rw@begin\fi\fi\fi}
\def\rw@next{\rwr@teswitch\n@xt}
\def\rw@@d{\backtotext} \let\rw@end=\relax
\let\backtotext=\relax

\newdimen\refindent     \refindent=30pt
\def\Textindent#1{\noindent\llap{#1\enspace}\ignorespaces}
\def\refitem#1{\par\hangafter=0 \hangindent=\refindent\Textindent{#1}}
\def\REFNUM#1{\space@ver{}\refch@ck\firstreflinetrue%
\global\advance\referencecount by 1 \xdef#1{\the\referencecount}}
\def\refnum#1{\space@ver{}\refch@ck\firstreflinetrue%
\global\advance\referencecount by 1\xdef#1{\the\referencecount}\refend}

\def\REF#1{\REFNUM#1%
\immediate\write\referencewrite{%
\noexpand\refitem{#1.}}%
\begingroup\obeyendofline\rw@start}
\def\ref{\refnum\?%
\immediate\write\referencewrite{\noexpand\refitem{\?.}}%
\begingroup\obeyendofline\rw@start}
\def\Ref#1{\refnum#1%
\immediate\write\referencewrite{\noexpand\refitem{#1.}}%
\begingroup\obeyendofline\rw@start}
\def\REFS#1{\REFNUM#1\global\lastrefsbegincount=\referencecount%
\immediate\write\referencewrite{\noexpand\refitem{#1.}}%
\begingroup\obeyendofline\rw@start}

\def\cite#1{\refmark#1}
\catcode`@=12 
%

\baselineskip=13pt plus 1pt minus 1pt

\proofmodefalse
\chapternumberstrue
\forwardreferencefalse
\ifproofmode
\immediate\openout2=allcrossreferfile \fi
\ifforwardreference\input labelfile
\ifproofmode\immediate\openout1=labelfile \fi\fi
\noblackboxes

\def\rd{{\rm d}}
\def\Ree{\mathop{\Re e}}
\def\wcp#1#2{\relax\ifmmode{\IP^{#2}_{#1}}\else $\IP^{#2}_{#1}$\fi}
%
\nopagenumbers\pageno=-1
\null\vskip-40pt
\rightline{\eightrm UTTG-16-95, HUB-IEP-95/23}\vskip-3pt
\rightline{\eightrm hepth/9511152}\vskip-3pt
\vskip .7truein
\centerline{\seventeenrm On Semi-Periods}
\bigskip\bigskip
\bigskip \bigskip
\centerline{{\csc A.C.~Avram}$^{1}$,\quad
            {\csc E.~Derrick}$^{2}$\footnote{$^{\natural}$}
                   {\eightrm Alexander von Humboldt Fellow},\quad
            {\csc D.~Jan\v{c}i\'{c}}$^{1}$}
\bigskip\bigskip
\centerline{
\vtop{\baselineskip=12pt\hsize = 2.0truein
\centerline{$^1$\it Theory Group}
\centerline{\it Department of Physics}
\centerline{\it University of Texas}
\centerline{\it Austin, TX 78712 U.S.A}}\quad
\vtop{\baselineskip=12pt\hsize = 2.0truein
\centerline{$^2$\it Humboldt Universit\"at zu Berlin}
\centerline{\it Institut f\"ur Physik}
\centerline{\it Invalidenstrasse 110}
\centerline{\it D-10115 Berlin, Germany}} }
\bigskip \bigskip
\vbox{\centerline{\bf ABSTRACT}
\vskip.2truein
\vbox{\baselineskip 12pt\noindent
The periods of the three--form  on a \cym\ are found as solutions
of the Picard--Fuchs equations;
however, the toric varietal method leads to
a generalized hypergeometric system of equations,
first introduced by Gelfand, Kapranov and Zelevinski,
which has more solutions than just the periods.
This same extended set of equations can be derived from
symmetry considerations.
Semi-periods are solutions of the extended GKZ system.
They are obtained by integration of
the three--form over chains;
these chains can be used to construct cycles which,
when integrated over, give periods.
In simple examples we are able to obtain the complete set of solutions for the
GKZ system. We also conjecture that a certain modification of the
method will generate the full space of solutions in general.}}
\vfill\eject
\headline={\ifproofmode\hfil\eightrm draft:\ 21 Nov 95\else\hfil\fi}
\pageno=1\footline={\rm\hfil\folio\hfil}
\section{intro}
{Preamble}
The moduli space of a \cym\ naturally splits into two
spaces:
the space $M_{21}$ of complex structure parameters, and
the space $M_{11}$ of (complexified) \K\ class parameters.
Calculations involving the complex structure parameters are
exact,
while those involving the \K\ class parameters are
corrected by instantons
(for a comprehensive review, see
  ~\Ref{\rBeast}{T.~H\"ubsch, {\it \cys ---A Bestiary for
       Physicists}\Z (World Scientific, Singapore, 1992).}).
Nevertheless,
under mirror symmetry, $M_{11}$ of one manifold is related
to $M_{21}$ of the mirror manifold.
Thus the parameters of the low energy effective action
of a string theory compactified on a \cym\
can be calculated exactly by studying the space $M_{21}$ of
the manifold, and the space $M_{21}$ of its mirror.

Since the discovery that Special Geometry applies to the
moduli space of \cys
  \REF{\rRolling}{P.~Candelas, P.S.~Green and T.~H\"ubsch,
      \npb{330} (1990) 49.}
  \REF{\rAS}{A.~Strominger, \cmp{133} (1990) 163.}
  \REF{\rCd}{P.~Candelas and X.~C.~de~la~Ossa, \npb{355} (1991) 455.}
 \ \cite{{\rRolling,\rAS,\rCd}},
the calculation of the periods
of the three-form has become something of an industry.
The space of complex structures is fully described by the
set of periods, in the sense that by knowing the periods,
one can calculate the metric on the moduli space and hence the kinetic term
and the Yukawa couplings in the low energy effective action.

Begin with the case of a family of hypersurfaces
${\cal M}_{\ph}$ defined as the zero-set of a defining polynomial $p_\ph$.
The periods are defined as
$$ \vp_j(\ph_\a)~~ \define ~~\int_{\g^j} \O(\ph_\a)~, \eqlabel{ePerDef}$$
where $\O(\ph_\a)$ is the nowhere vanishing holomorphic
three-form
{}~\REF{\rPC}{P.~Candelas, \npb{298} (1988) 458.}
\REF{\rMA}{M.~Atiyah, R.~Bott and L.~G{\aa}rding,
                       Acta Math. {\bf 131} (1973) 145.}
\cite{{\rPC,\rMA}}

$$ \O(\ph_\a)~~ \define
           ~~{\rm Res}_{{\cal M}_\ph}\bigg[{(x\rd^4x)\over p_\ph}\bigg] $$
on the Calabi-Yau three-fold specified by the parameters $\ph_\a$,
and $\g^j$ is a three-cycle labelled by $j$.
The four-differential
$(x\rd^4x)$ is the `natural' one: on a weighted projective $N$-space
 $\wcp{(k_1,\ldots,k_{N+1})}{N}$, we have
$$ (x\rd^Nx)~~ \define
             ~~{1\over (N{+}1)!} \e^{i_1 i_2\cdots i_{N+1}}
                k_{i_1} x_{i_1} \rd x_{i_2} \cdots \rd x_{i_{N+1}}~, $$
where $k_i$ are the weights of the coordinates. That is, in
 $\wcp{(k_1,\ldots,k_{N+1})}{N}$,
$$ (x_1,\ldots,x_{N+1}) \cong (\l^{k_1}x_1,\ldots,\l^{k_{N+1}}x_{N+1})~,
             \qquad \l \in \IC^*~. $$
The period~\eqref{ePerDef}\ will here be calculated by choosing one of the
standard coordinate patches in $\wcp{(k_1,\ldots,k_5)}{4}$, ${\cal U}_m$,
where $x_m\neq0$ so that $x_m=1$ by projectivity. There
{}~\Ref{\rSpokes}{P.~Berglund, E.~Derrick, T.~H\"ubsch and
    D.~Jan\v{c}i\'{c}, \npb{420} (1994) 268, hep-th/9311143.}:
$$ \vp^{(m)}_j(\ph_\a)~~ = ~~C \int_{\G^j}
          {\prod_{i\neq m}\rd x_i\over p_\ph\vert_m}~,
\eqlabel{eOM}$$
with $C$ a convenient prefactor. This is easily seen to apply
for Calabi-Yau weighted hypersurfaces of arbitrary dimension.

We separate the polynomial into a reference polynomial $p_0$
independent of the moduli~$\ph$, and a perturbative part $\D$
which does depend on the $\ph$.
This amounts to choosing a reference point in moduli space,
and expanding around that point.
The basic idea here is,
while working in the patch ${\cal U}_m$ but suppressing the label,
to expand $1/p_\ph$ around $1/p_0$, and utilize the Laplace transform
$$ {1\over p_\ph}~ = ~
{1\over p_0 + \D} ~=~
    \sum_{n=0}^\infty {(-\D)^n \over \- (p_0)^{n+1} }
{}~=~ {d\over k_m} \int_0^\infty \rd x_m \sum_n {(-\D)^n \over n!} e^{- p_0}
{}~,\quad \Ree(p_0)>0~,\eqlabel{eLaplace}$$
which produces a ``small-$\phi$''
expansion of the periods~\eqref{eOM}.
We have relabelled the Laplace transform parameter
to $x_m^{d/k_m}$, and so the expansion appears as if in homogeneous
coordinates.
With a choice of the poly-contours $\{\G^j\}$, Eq.~\eqref{eOM}\ may be
considered a definition of the periods.

The semi-period is a building block for a period.
In other words, it is the integral of the three--form
over a chain instead of a cycle, with the understanding that
the chain can be manipulated in some way to build cycles.
The semi-period construction we will discuss is
naturally related to the period construction in
  \cite{\rSpokes}.
The prototypical semi-period we write as
$$ F_0 = \int_V \, \rd^{N+1} x\, \sum_{n=0}^\infty\, {(-\D)^n \over n!}\,
                              e^{-p_0}~,
\eqlabel{protosp}$$
where $V$ is the ${N+1}$-chain
$\{x_1,\ldots,x_{N+1}$ real and positive$\}$.

As was worked out in \cite{\rSpokes}, these chains are enough
to build cycles and thus to calculate periods.
The advantage of this approach is its ease of calculation of a
full set of periods in different regions of moduli space.
This method lends itself to many different constructions:
all the types of polynomial hypersurfaces discussed in
  \REF{\rCfC}{P.~Berglund and T.~H\"ubsch,
              \npb{411} (1994) 223, hep-th/9303158.}
  \REF{\rCLS}{P.~Candelas, M.~Lynker and R.~Schimmrigk,
                      \npb{341} (1990) 383.}\
  \cite{{\rCfC,\rCLS}},
(weighted) complete intersection spaces
  \REF{\rCYCI}{T.~H\"ubsch, \cmp{108} (1987) 291.}
  \REF{\rGH}{P.~Green and T.~H\"ubsch, \cmp{109} (1987) 99.}
  \REF{\rCDLS}{P.~Candelas, A.M.~Dale, C.~A.~L\"utken and R.~Schimmrigk,\Z
       \npb{298} (1988) 493.}
  \REF{\rGRY}{B.~Greene, S.-S.~Roan and S.-T.~Yau, \cmp{142} (1991) 245.}\
\cite{{\rBeast, \rCYCI, \rGH, \rCDLS, \rGRY}},
generalized Calabi-Yau manifolds
 \REF{\rZ}{P.~Candelas, E.~Derrick and L.~Parkes,
        \npb{407} (1993) 115,\Z hep-th/9304045.}
  \REF{\rRolf}{R.~Schimmrigk, \prl{70} (1993) 3688, hep-th/9210062.}\
\cite{{\rZ, \rRolf}},
Landau--Ginzburg vacua
  \REF{\rMaxSkII}{M.~Kreuzer and H.~Skarke, ``On the Classification
       of Quasihomogeneous Functions'', CERN preprint CERN-TH-6373/92.}
  \REF{\rMaxSkI}{M.~Kreuzer and H.~Skarke, \npb{388} (1992) 113,
               hep-th/9205004.}
  \REF{\rAR}{A.~Klemm and R.~Schimmrigk,
       \npb{411} (1994) 559, hep-th/9204060.}\
\cite{{\rMaxSkII,\rMaxSkI,\rAR}},
and toric varietal constructions related to these and other types
 ~\REF{\rBatyrev}{V.~Batyrev,
    Duke Math. Journal, Vol 69, No 2, (1993) 349.}
  \REF{\rHKTY}{S.~Hosono, A.~Klemm, S.~Theisen and S.-T.~Yau,
     \cmp{167} (1995) 301, hep-th/9308122.}
  \REF{\rHKTYII}{S.~Hosono, A.~Klemm, S.~Theisen and S.-T.~Yau,
      \npb{433} (1995) 501, hep-th/9406055.}
  \REF{\rCdK}{P.~Candelas, X. de la Ossa and S. Katz,
    \npb{450} (1995) 267, hep-th/9412117.}\
\cite{{\rBatyrev,\rHKTY,\rHKTYII,\rCdK}}.

A semi-period is also the solution of
a generalized hypergeometric system of equations.
The periods are solutions of the Picard-Fuchs (PF) equations.
In the toric varietal approach,
differential equations are constructed based on
the points in the dual polyhedron and the generators of Mori cone
  ~\REF{\rBKK}{P.~Berglund, S.~Katz and  A.~Klemm,
    \npb{456} (1995) 153, hep-th/9506091.}
\cite{{\rBatyrev,\rHKTY,\rBKK}}.
The system of equations obtained in this manner has extra solutions beside
periods, since the complete set of solutions is larger than
the set of linearly independent periods.
We show by construction that semi-periods such as \eqref{protosp}
are such extra solutions.
All of the periods can be obtained either as the solutions to PF equations,
or as linear combinations of semi-periods.

In many if not all cases the same generalized hypergeometric system that was
obtained from toric data can be constructed based on the symmetries of the
period, where the period is expressed as an integral over a cycle
 ~\REF{\rY}{S.-T. Yau, talk presented at Workshop on Mirror Symmetry
    and S-Duality, ICTP, Trieste, Italy, June 1995.}
\cite{{\rHKTY,\rY}}.
These equations are also satisfied by a semi-period,
which is expressed using the
same integrand, but integrated over a particular chain, not a cycle.
Not every chain yields a solution to the differential equations;
for those that do, we call the integral a semi-period.

Our aim in this paper is to construct, calculate and
investigate the semi-period solutions to the $\Delta^*$
hypergeometric system of equations.
The layout of the paper is as follows:
Section~\chapref{ghs} describes the construction of
the generalized hypergeometric system of equations using the
toric varietal approach and
Section~\chapref{sex} illustrates the procedure
by doing a simple example: the $\IZ_3$ torus.
An interesting set of examples are explored in
Sections~\chapref{example} and \chapref{exampleIII}.

\section{ghs}
{Construction of the $\Delta^*$ Hypergeometric System
 associated to Toric Manifolds}
In this section we briefly review the construction of the
$\Delta^*$
hypergeometric system for {\cys } described as the zero loci of
homogeneous polynomials in weighted projective spaces.

We denote a weighted projective space as $\IP^r_{\bf k}$,
by which we mean the set of complex $x_i$ (not all zero)
identified under
$(x_1, x_2, \ldots, x_{r+1}) \sim
  (\l^{k_1} x_1,\l^{k_2} x_2, \ldots, \l^{k_{r+1}} x_{r+1}) $
for all $\l\neq 0$.
Consider the \cym\ $\IP^r_{\bf k}[d]$ with $\sum_{i=1}^{r+1}
k_{i} = d$, where  $d$ is the degree of the defining
polynomial whose general expression is :
$$
p = \sum_{\bf m} c_{\bf m} x^{\bf m}
$$
Here $\bf m$ is the degree vector such that ${\bf m} \cdot {\bf k} = d$.

We associate to each monomial a point with coordinates
${\bf m} =(m_{1}, \dots , m_{r+1})$ in the lattice $\IZ^{r+1}$.
The set of all monomials compatible with ${\bf m} \cdot {\bf k} =d$ will
define a convex polyhedron $\Delta$ in $\IZ^{r+1} \otimes \IR$, which is
moreover reflexive (in the sense of \cite{\rBatyrev}).
This polyhedron lies in a hyperplane defined by the weight vector normal to it
and will contain all the points of the sublattice that lie within its
boundaries.
By uniting the vertices of the polyhedron $\Delta$ with the origin
we obtain a
cone whose dual can be easily constructed by the method described in
\cite{\rBatyrev}.

The vertices of the dual polyhedron $\Delta^*$ are given by the first
intersection point between the generators of the dual
cone and the dual lattice.
Because none of the weights in the definition of
$\IP^r_{\bf k}$ is zero, we
can project the dual polyhedron on a
$\IZ^{r} \otimes \IR$ subspace and then find a basis for the sublattice
defined by the projected points of the dual polyhedron.
We take the unique interior point as the origin of the new coordinate system,
and then put back the dual polyhedron in
$\IZ^{r+1} \otimes \IR$ along the hyperplane $x_{0}=1$.

This leaves us with the following set of dual points:
$$
 \bar{\nu}_{i}^{*} = (1, \nu_{i}^{* 1}, \ldots , \nu_{i}^{* r})
\qquad i=1, \ldots ,p   ~~~,
\eqlabel{nu2}
$$
and the interior point
$$
 \bar{\nu}_{0}^{*} = (1,0, \ldots, 0)~.
$$

The space of linear dependence relations between the dual points,
$\sum_{i=0}^p \, l_{i}^{a} \, \bar{ \nu}_{i}^{*} = 0$,
will have dimension $p-r$. Nevertheless as Batyrev remarked in \cite{\rBatyrev}
we are interested only
in points that do not lie in
the interior of codimension $1$ faces. Assuming there are $n +1$
of them, we denote $A = \{ \bar{\nu}_{0}^{*},  \bar{\nu}_{i_1}^{*},
\dots, \bar{\nu}_{i_n}^{*} \}$ the subset of such points in the dual
polyhedron.
The linear dependences between points in $A$ are described by the lattice of
rank $n - r$:

$$
 L = \{ (l_{0}, \dots, l_{n}) \in \IZ^{n+1} \mid \sum_{i=0}^n\, l_{i}^{a} \,
\bar{ \nu}_{i}^{*} = 0, \bar{ \nu}_{i}^{*} \in A \}
\eqlabel{lindep}
$$
Considering the affine complex space $\IC^{p+1}$ with coordinates $(a_{0},
\dots, a_{p})$ we define a consistent system of
differential operators:
$$\eqalign{
 Z_j \, &= \, \sum_{i=0}^{n} \, \bar{ \nu}^*_{i,j} \, a_i
{\partial\over\partial a_i} \, - \, \beta_j  \cr
 D_{{\bf l}^a} \, &= \, \prod_{l_i>0} \left({\partial \over \partial
a_i}\right)^{l_i^a} \, - \, \prod_{l_i<0} \left({\partial \over \partial
a_i}\right)^{-l_i^a} ; \quad  {\bf l}^a \in L \cr}
\eqlabel{sys}
$$
The exponent of $\bf\beta$ that appears in the definition of the differential
operators $Z_j$ will be taken
to be $(-1, 0, \ldots, 0)$.
With this choice the period integrals
vanish when acted upon by the operators  \eqref{sys}  \cite{\rBatyrev}.

The set of operators \eqref{sys} taken together form the
$\Delta^*$ hypergeometric system of equations.
This is an example of a class of generalized hypergeometric equations
described by
Gel'fand, Kapranov and Zelevinski \
  \Ref{\rgelfand}{I.~M.~Gelfand, A.~V.~Zelevinski and M.~M.~Kapranov,
   Func. Anal. Appl. Vol. 23 No. 2 (1989) 94.},
and hence is also called a GKZ system.
In general,
the $\Delta^*$ system is not enough to determine the Picard-Fuchs equations;
it must be extended by supplementing further differential
operators.  The methods for doing this are discussed in \
\REF{\rHLY}{S.~Hosono, B.~H.~Lian and S.-T.~Yau,
     ``GKZ-Generalized Hypergeometric System in Mirror Symmetry
      of Calabi-Yau Hypersurfaces'', alg-geom/9511001.}
\cite{{\rHKTY, \rBKK, \rHLY}}.
In other words, there are more solutions to this system of equations than
just the periods.
We conjecture that semi-periods form a complete set of solutions.

The periods and semi-periods
depend on  $a_i$ through $\zeta^a$ \cite{\rHKTY}, defined as:
$$
\zeta^a = (-1)^{{ \hat l}_0^a} \, {\bf a}^{{\bf{\hat l}}^a} ~.\eqlabel{zetadef}
$$
We used ${\bf {\hat l}}^a$ to denote the generators of the Mori cone which can
be found by standard
methods(see for example \cite{\rBKK}).
The variables $\z^a$
are not only the natural choices in order to
satisfy the linear constraints imposed by the $Z^j$,
but are also good coordinates on the
moduli space for describing the large complex structure limit of the mirror
manifold.

\section{sex}{Simple Example}
\subsection{The $\Delta^*$ Hypergeometric System for the $\IZ_3$ Torus}
The $\IZ_3$ torus can be described as a cubic hypersurface in $\IP^2$.
The  vector of weights is ${\bf k} = (1,1,1)$,
and so the vertices of the dual polyhedron become
$$
 \bar{\nu}^*_1 = (1,-1,-1) \quad \bar{\nu}^*_2 = (1,1,0) \quad
\bar{\nu}^*_3 = (1,0,1) ~, $$
and the interior point is $ \bar{\nu}^*_0 = (1,0,0)$.

There is one relation of the form
$ \sum_i l_i \bar{\nu}^*_i = 0$,
satisfied by ${\bf l} = (-3,1,1,1)$.
Taking this into the account, the differential operators defined by \eqref{sys}
will be of the following form:
$$ D = {\partial\over\partial a_1}\, {\partial\over\partial a_2}\,
       {\partial\over\partial a_3}
        - \left({\partial\over\partial a_0}\right)^{3}
\eqlabel{Dtorus}$$
and
$$\eqalign{
 Z_1  &= \left( \sum_{i=0}^3 a_i {\partial\over\partial a_i} +1
            \right)  \cr
 Z_2  &= \left(  a_2 {\partial\over\partial a_2} -
                    a_1 {\partial\over\partial a_1}    \right)  \cr
 Z_3  &= \left(  a_3 {\partial\over\partial a_3} -
                    a_1 {\partial\over\partial a_1}    \right) ~. \cr}
\eqlabel{Zitorus}$$
The equations $Z_i \Pi = 0$ are solved by $\Pi$ of the form
$$\Pi = {1\over a_0} \tilde\Pi\left(-{a_1 a_2 a_3 \over a_0^3}\right)~,$$
and the equation $D \Pi = 0 $ becomes
$$ \left[ \Theta^3 -\z (3\Theta + 3)(3\Theta +2)(3\Theta + 1)\right]
   \tilde\Pi = 0~,\eqlabel{Dtorus}$$
where $\Theta = \z {\partial\over\partial \z}$, and $\z$ is the argument
of $\tilde\Pi$, found from \eqref{zetadef}.

Solving the generalized hypergeometric equation one has
{}~\Ref\rErdelyi{A.~Erd\'elyi, F.~Oberhettinger, W.~Magnus and
       F.~G.~Tricomi, {\it Higher Transcendental Functions, 3 Vols.}
       (McGraw--Hill, New York, 1953).}\ :
$$\tilde \Pi = \Psymbol{0&1/3&0\cr 0&2/3&1\cr 0&1&0\cr}
                       {\cr \cr 3^3 \z \cr \cr}
{}~.$$
We will be interested in the region where $a_0$ is small.
Since $\z$ is related to $a_0^{-3}$, we should look at the solutions
near $\z\rightarrow\infty$.
Remembering that $\Pi = {1\over a_0} \tilde\Pi$, and setting
$a_1 = a_2 = a_3 = 1$, $\z = -1/a_0^3$, the solutions to the
$\Delta^*$-hypergeometric system are
$$\eqalign{
  \Pi_0 &=\hphantom{a_0^2}\
           {}_2F_1\left({1\over 3},{1\over 3};{2\over 3};
                        -\left({a_0\over 3}\right)^3\right) \cr
  \Pi_1 &= a_0\  {}_2F_1\left({2\over 3},{2\over 3};{4\over 3};
                        -\left({a_0\over 3}\right)^3\right) \cr
  \Pi_2 &= a^2_0\ {}_3F_2\left(1,1,1;{4\over 3},{5\over 3};
                        -\left({a_0\over 3}\right)^3\right)~. \cr}
\eqlabel{Pistorus}$$

\subsection{The Equation from Symmetry Considerations}
Suppose we look at the period defined as
$$ \Pi = \int_\G {x_1 \rd x_2 \rd x_3 \over
 a_1 x_1^3 +a_2 x_2^3 +a_3 x_3^3 + a_0 \ x_1 x_2 x_3 }~,
\eqlabel{perdef}$$
where $\G$ is some cycle.  Since $\Pi$ is
independent of $x_1$,
it is equivalent to
$$ \Pi = {1\over 2 \pi i} \int_{\G'} {\rd x_1 \rd x_2 \rd x_3 \over
 a_1 x_1^3 +a_2 x_2^3 +a_3 x_3^3 + a_0 \ x_1 x_2 x_3 }~,
\eqlabel{perdefII}$$
where $\G' = C_1 \times \G$, and $C_1$ is a loop around $x_1=0$.
Now look at the symmetries of $\Pi$ as a function of the $a_i$:
$$
\Pi(a_0,\l^3 a_1, \l^{-3} a_2,a_3) = \Pi(a_0,a_1,a_2,a_3) ~,
 \eqlabel{symtorI}$$
and the same with $a_2\rightarrow a_3$, so
$$\eqalign{
 a_1 \pd{\Pi}{a_1}  &= a_2 \pd{\Pi}{a_2} \cr
 a_1 \pd{\Pi}{a_1}  &= a_3 \pd{\Pi}{a_3} ~,\cr}\eqlabel{ZsymtorI}$$
while
$$\Pi(\l^3a_0,\l^3 a_1,\l^3 a_2,\l^3 a_3) =\l^{-3} \Pi(a_0,a_1,a_2,a_3)
\eqlabel{symtorII}$$
implies that
$$ a_0 \pd{\Pi}{a_0}  +\sum_{i=1}^3 a_i \pd{\Pi}{a_i} = -\Pi~.
\eqlabel{ZsymtorII}$$
These are the $Z_i$ equations \eqref{Zitorus}.
Finally, since $x_1^3 x_2^3 x_3^3 = (x_1 x_2 x_3)^3$,
the period satisfies the relation
$$ \pd{}{a_1} \pd{}{a_2} \pd{}{a_3} \Pi=
   \left( {\partial\over\partial a_0}\right)^3 \Pi ~,
\eqlabel{Dsymtor}$$
which is the $D$ equation \eqref{Dtorus}.

Thus we are able to find the equations from consideration
of the symmetries of the period,
without going through the toric variety construction.

\subsection{The Semi-Periods for the $\IZ_3$ Torus}
Consider the integrand in the definition of the period \eqref{perdef},
and note that we can take a Laplace transform as long as $\Ree(p)>0$:
$$ {x_1 \rd x_2 \rd x_3 \over p}  =
   \int_{x_1=0}^\infty \rd^3 x\  e^{-p}~.$$
Taking the polynomial to be
$$p =  a_1 x_1^3 +a_2 x_2^3 +a_3 x_3^3 + a_0 x_1 x_2 x_3 ~,$$
we define the integration contour
$ V = \{x_1, x_2, x_3 \hbox{ real and positive} \} $
and construct
$$ F = \int_V \rd^3 x\  e^{-p} ~.\eqlabel{FZ}$$
$F$ as defined also satisfies the relations \eqref{ZsymtorI}
\eqref{ZsymtorII} \eqref{Dsymtor} derived
from symmetry considerations above,
and hence is the extra solution besides the periods.
This is a semi-period.
By using the symmetries \eqref{symtorI} \eqref{symtorII},
we bring the polynomial in the integrand of $F$ to canonical form
$$p =   x_1^3 + x_2^3 + x_3^3 + \phi\  x_1 x_2 x_3 ~.$$
Now it is easy to calculate $F(\phi)$:
$$\eqalign{
F &= \sum_{n=0}^\infty {(-\phi)^n\over n!} \
    \prod_{i=1}^3 \int_0^\infty \rd x_i \, x_i^n\, e^{-x_i^3}\cr
  &= {1\over 3^2} \sum_{n=0}^\infty
     {(-\phi)^n\over n!} \, \Gamma^3\left( {n+1\over 3}\right) \cr}
\eqlabel{FZcalc}$$
which is, of course, a linear combination of
$\Pi_0$, $\Pi_1$ and $\Pi_2$ of \eqref{Pistorus}, with $a_0 = \phi$.

But now, consider integrating over a slightly different contour.
Let $x_2$ run from 0 to $\o \infty$, $\o$ being a cube root of unity,
and call this new contour $A_2 V$.
This still satisfies the equations, but by a change of variables
we see that
$$ \int_{A_2 V} \rd^3 x\  e^{-p(\phi)} = \o F(\o \phi) ~.$$
There are three semi-periods we can form in this way,
summarized by integrating
$x_2$ from 0 to $\o^{\d} \infty$, $\d=0,1,2$.
This inserts a factor of $\o^{n+1}$ in the summation of
\eqref{FZcalc}, and leads to
three independent linear combinations of \eqref{Pistorus}.

So what happens when we form a spoke in the $x_2$ plane,
that is, we integrate $x_2$ along a contour that comes in infinity
along one of these paths, and returns along the other?
The integration contour is $V - A_2 V$, so the result of
the integration is $F(\phi) - \o F(\o \phi)$,
which is a linear combination of $\Pi_0$ and $\Pi_1$.
Note that in generating periods for the torus by using spokes
\cite{\rSpokes},
we take exactly these sorts of contours, and
the periods are linear combinations of
$\Pi_0$ and $\Pi_1$.
Details of this example are in the Appendix in \cite{\rZ}.
A cycle could be generated, for example, by integrating over
$ (1-A_2) (1-A_3) V$, and the period that results is
$ F(\phi) - 2 \o F(\o \phi) + \o^2 F(\o^2 \phi)$.

Why do the semi-periods satisfy the $\Delta^*$ hypergeometric
system?
The differential equations can be generated by considering the
symmetries of the period  integral \eqref{perdefII}.
The symmetry in \eqref{symtorI} is demonstrated by a change of variables
in the integrand of \eqref{perdefII},
$x_1\rightarrow \l x_1$ and $x_2\rightarrow \l^{-1} x_2$.
Since the region of integration has no boundary,
the expression is the same.
Applying the same consideration to \eqref{FZ} or \eqref{FZcalc},
we see that (taking $\l$ to be real), the boundary is unchanged
and again, the expression is the same.
Thus the same symmetries are exhibited by the semi-periods,
since the boundary of the chain has been chosen properly.

The $\Delta^*$ system contains the Picard-Fuchs equations, and is thus
solved by the periods;
but it is also solved by the semi-periods.
The methods for reducing the order of the $\Delta^*$ system
to the order of the PF equations are discussed in
\cite{{\rHKTY,\rBKK,\rHLY}}.
In~\cite{\rBKK}, use is made of further properties of
the periods that can be derived using partial integration.
These properties are not shared by the semi-periods,
and hence this method naturally selects periods over semi-periods.
References~\cite{\rHKTY} and~\cite{\rHLY} find other ways
of selecting periods, either by reducing the order of the
$\Delta^*$ system, or imposing additional differential equations.

There are, of course,  more semi-periods than periods.
Even though we have not been able to explicitly construct them in
general, we believe that all solutions of the $\Delta^*$ system can be
obtained by integrating the three--form over well chosen chains.
We have not found a general principle that guarantees that
all solutions to the $\Delta^*$ system can be found as
integrals over chains ---
it is, however, true in simple cases.
It seems likely that
the selection of these chains is highly model dependent.
For the types of chains we consider,
the number of semi-periods generated in a particular corner of
moduli space depends on the symmetries of $p_0$,
the reference polynomial.

\section{example}{A Known Example}
Here is an example for which the periods are discussed
in several places,
  \REF{\rGoryVII}{P.~Berglund, P.~Candelas, X.~de la Ossa, A.~Font,
       T.~H\"ubsch, D.~Jan\v{c}i\'{c} and F.~Quevedo,
        \npb{419} (1994) 352, hep-th/9308005.}\
namely Section 4.2 of \cite{\rGoryVII},
Section 3.1 of \cite{\rSpokes} and
Section 3.1 of \cite{\rBKK}.
Let us examine the semi-periods for a particular family of
hypersurfaces, $\cp{(2,2,3,7,7)}[21]^{50,11}$,
where the superscripts indicate the
Hodge numbers $b_{2,1}=50$ and $b_{1,1}=11$,
with the polynomial chosen as
$$ p = a_1 y_1^7 y_4 + a_2 y_2^7 y_5 + a_3 y_3^7 +
       a_4 y_4^3 + a_5 y_5^3 + a_0 y_1 y_2 y_3 y_4 y_5 +
       a_6 (y_1 y_2 y_3)^3 ~.$$
To calculate a semi-period $F$, we must
pick a corner of moduli space.
Supposing we treat the last two terms in $p$ as
perturbations, $F$ is  calculated to be
$$\eqalign{
 F
   &=
  \sum_{m,k=0}^{\infty} \,{(-a_0)^m \over m!}{(-a_6)^k \over k!}\cr
   &\quad\quad\quad\times
  \int_0^\infty \rd^5 y \ (y_1 y_2 y_3)^{m+3k}\, (y_4 y_5)^m\
  e^{-\left(a_1 y_1^7 y_4 + a_2 y_2^7 y_5 + a_3 y_3^7 +
       a_4 y_4^3 + a_5 y_5^3 \right)} \cr
  &= {1\over 3^2 7^3 }\,
  \sum_{{m,k=0 \atop 2m+2>k}}^{\infty}
             \,{(-a_0)^m \over m!}{(-a_6)^k \over k!}\,
  {\Gamma^3\left({m+3k+1\over 7}\right)\over
         (a_1 a_2 a_3)^{m+3k+1\over 7} }\
  {\Gamma^2\left({2m-k+2\over 7}\right)\over
         (a_4 a_5)^{2m-k+2\over 7} }        ~.       \cr}
\eqlabel{F}$$

Things to notice about this in reference to the references
\cite{{\rGoryVII,\rSpokes,\rBKK}}:
\item{$\bullet$}This can be expressed as a function of the
two large complex structure parameters  as defined in
\cite{{\rGoryVII,\rBKK}},
$\zeta_1 = -{a_4 a_5 a_6\over a_0^3}$ and
$\zeta_2 = - {a_1 a_2 a_3 \over a_0 a_6^2}$, via
$$ \eqalign{
F &= {1\over a_0 }\tilde F(\zeta_1,\zeta_2)\cr
\tilde F &= \sum_{r=0}^1 \sum_{j,l=0}^\infty \,
             \left(-{1\over \zeta_1}\right)^{{2j+2-r\over 7}}\,
             \left(-{1\over \zeta_2}\right)^{l+{j+1+3r\over 7}}\,
        { (-1)^{l+j+r}\over \Gamma(l+j+1)\Gamma(2l+1+r) }\cr
&\quad\quad\quad\times
  \Gamma^3\left(l+{j+1+3r\over 7}\right)\,
  \Gamma^2\left({2j+2-r\over 7}\right)~.\cr}
\eqlabel{FII}$$
$F$ has been calculated in an expansion around large $\z_1$
and $\z_2$.
\item{$\bullet$}This obeys the equations found in \cite{\rBKK},
$$\eqalign{
  \partial^3_{a_0} F -
        \partial_{a_4}\partial_{a_5}\partial_{a_6} F &=0\cr
  \partial^2_{a_6} \partial_{a_0} F -
        \partial_{a_1}\partial_{a_2}\partial_{a_3} F &=0~,
\cr}\eqlabel{pfexample}$$
which become the extended Picard--Fuchs equations
$$\eqalign{
  \Big\{& \Theta_1^2 (\Theta_1 - 2 \Theta_2) -
   \z_1 (3 \Theta_1 + \Theta_2 + 3)\, (3 \Theta_1 + \Theta_2 + 2)\,
          (3 \Theta_1 + \Theta_2 + 1) \Big\} \tilde F = 0\cr
  \Big\{& \Theta_2^3 -
   \z_2 ( \Theta_1 -2 \Theta_2 -1)\, ( \Theta_1 -2 \Theta_2)\,
          (3 \Theta_1 + \Theta_2 + 1) \Big\} \tilde F = 0~,\cr}$$
using $\Theta_i =\z_i \partial_{\z_i}$.
\item{$\bullet$}
There are seven semi-periods that are easy to find in this expansion.
These were exploited in \cite{\rSpokes} to calculate the
periods.  The different semi-periods are obtained by
integrating over different chains in \eqref{F},
as detailed in \cite{\rSpokes}.
Consider changing the region of integration in \eqref{F} so that
$y_1$ runs from 0 to $\l \infty$,
where $\l$ is a seventh root of unity.
The net effect is to change $F$ by
$$ A : F(a_0,a_6) \rightarrow \l F(\l a_0, \l^3 a_6) ~.$$
Explicitly, this leads to the insertion of $\l^{m+3k+1}$
inside the summation in \eqref{F}, or an insertion of $\l^{j+1+3r}$
inside the summation in \eqref{FII}.
This leads to seven semi-periods.
It has been shown in \cite{\rSpokes}
that the six independent periods can be calculated using spokes,
that is, using differences of semi-periods generated in
this way.
The action of $A$ is discussed in \cite{{\rGoryVII,\rSpokes}}.
\section{exampleIII}
{Nesting Example}
The following example is interesting because it
shows how $K3$ and $T^2$ can appear.
In this section
we extend an argument by Klemm, Lerche and Mayr \
  \Ref{\rKLM}{A.~Klemm, W.~Lerche and P.~Mayr, 
    \plb{357} (1995) 313, hep-th/9506112.} \
to include the semi-periods;
see also \
   \Ref{\rLYII}{B.~H.~Lian and S.-T.~Yau, ``Mirror Maps, Modular Relations
      and Hypergeometric Series II'', hep-th/9507153.} \
and \cite{\rHKTY}.
Consider the manifold $\cp{(1,1,2,4,4)}[12]^{101,5}$,
and choose  a particular family of  hypersurfaces,
$$ p = a_1 x_1^{12} + a_2 x_2^{12} + a_3 x_3^{6} +
       a_4 x_4^3 + a_5 x_5^3 + a_0 x_1 x_2 x_3 x_4 x_5 +
       a_6 (x_1 x_2)^{6} + a_7 (x_1 x_2 x_3)^3 ~.$$
This is a $K3$ fibration, as can be seen by making
the substitutions $x_2 = \l x_1$, $\tilde x_1 = x_1^2$,
which leads to
$$ p = \tilde x_1^{6} (a_1 + \l^{12} a_2 + \l^{6} a_6)
        +a_3 x_3^{6} +   a_4 x_4^3 + a_5 x_5^3 + \l a_0 \tilde x_1 x_3 x_4 x_5
+
         + \l^3 a_7 (\tilde x_1 x_3)^3 ~.$$
The hypersurface $p=0$ describes a  $K_3$ hypersurface
\cp{(1,1,2,2)}[6].
By a similar substitution,
this $K3$ is shown to be an elliptic fibration over $\IP^1$ with generic
fiber \cp{(1,1,1)}[3], otherwise known as the $\IZ_3$ torus, described in
Section~\chapref{sex}.

The dual polyhedron of $\cp{(1,1,2,4,4)}[12]$ contains 8 points:
$$\eqalign{
 \bar{\nu}_0^* &= (1,\- 0,\- 0,\- 0,\- 0)\cr
 \bar{\nu}_1^* &= (1,\- 4,\- 0,\- 2,\- 3)\cr
 \bar{\nu}_2^* &= (1,\- 0,\- 0,\- 0,\- 1)\cr
 \bar{\nu}_3^* &= (1,\- 0,\- 0,\- 1,\- 0)\cr
 \bar{\nu}_4^* &= (1,\- 0,\- 1,\- 0,\- 0)\cr
 \bar{\nu}_5^* &= (1, - 1, - 1, - 1, - 1)\cr
 \bar{\nu}_6^* &= (1,\- 2,\- 0,\- 1,\- 2)\cr
 \bar{\nu}_7^* &= (1,\- 1,\- 0,\- 1,\- 1)\cr} $$
We find $6$ vectors in the Mori cone, which can all be expressed with
positive integer coefficients in terms of the following basis:
$$\eqalign{
{\bf l}^{1} &= ( - 3,\- 0,\- 0,\- 0,\- 1,\- 1,\- 0,\- 1)\cr
{\bf l}^{2} &= (\- 0,\- 0,\- 0,\- 1,\- 0,\- 0,\- 1, - 2)\cr
{\bf l}^{3} &= (\- 0,\- 1,\- 1,\- 0,\- 0,\- 0, - 2,\- 0)\cr
}
$$
The large complex structure coordinates are:
$$\eqalign{
\zeta_1 &= - {a_4 a_5 a_7 \over a_0^3} \cr
\zeta_2 &= {a_3 a_6 \over a_7^2}  \cr
\zeta_3 &= {a_1 a_2 \over a_6^2} \cr}
$$
In this particular example
the constraint operators $Z_j$ have the expressions:
$$\eqalign{
Z_1 &= \sum_{i=0}^{7} \, a_i {\partial\over\partial a_i} +1 \cr
Z_2 &= 4 a_1 {\partial\over\partial a_1}
       - a_5 {\partial\over\partial a_5}
      +2 a_6 {\partial\over\partial a_6}
      +  a_7 {\partial\over\partial a_7} \cr
Z_3 &=  a_4 {\partial\over\partial a_4}
       - a_5 {\partial\over\partial a_5} \cr
Z_4 &= 2 a_1 {\partial\over\partial a_1}
       + a_3 {\partial\over\partial a_3}
      - a_5 {\partial\over\partial a_5}
      +  a_6 {\partial\over\partial a_6}
      +  a_7 {\partial\over\partial a_7} \cr
Z_5 &= 3 a_1 {\partial\over\partial a_1}
       + a_2 {\partial\over\partial a_2}
       - a_5 {\partial\over\partial a_5}
      +2 a_6 {\partial\over\partial a_6}
      +  a_7 {\partial\over\partial a_7} \cr}
$$
The $D_{\bf l}$ operators are:
$$\eqalign{
D_1 &= {\partial\over\partial a_1} {\partial\over\partial a_2}
      -\left({\partial\over\partial a_6} \right)^2 \cr
D_2 &= {\partial\over\partial a_4} {\partial\over\partial a_5}
       {\partial\over\partial a_7}
       -\left({\partial\over\partial a_0} \right)^3 \cr
D_3 &= {\partial\over\partial a_3} {\partial\over\partial a_6}
      -\left({\partial\over\partial a_7} \right)^2 \cr}
$$
If we parametrize the solutions of the above system as follows:
$$
 F({\bf a})= {1 \over a_0} {\tilde F}(\zeta_i)
$$
then $\tilde{F}( \zeta_i)$ is a solution of $D_i\, \tilde F = 0$, with
$$\eqalign{
{\cal D}_1 &= \theta_{3}^2 - \zeta_{3} \prod_{i=0}^{1} (\theta_{2} - 2
\theta_{3}-i) \cr
{\cal D}_2 &= \theta_{1}^2 (\theta_{1}-2 \theta_{2}) - 3^3 \zeta_{1}
\prod_{i=1}^{3} (\theta_{1} +i/3) \cr
{\cal D}_3 &= \theta_{2} (\theta_{2}- 2 \theta_{3}) - \zeta_{2}
\prod_{i=0}^{1} (\theta_{1} -2 \theta_{2} -i) \cr}
\eqlabel{d3}
$$
where $ \theta_i = \zeta_i {\partial \over \partial \zeta_i}$.

The dimension of the space of solutions for the $\Delta^*$-generalized
hypergeometric system is finite and can be calculated by two methods.

The first method uses the associated indicial equations.
In the corner of the moduli space where all the $\zeta_i$'s are large,
we obtain the following system of indicial equations (after the change
of variables $\zeta_i = {1 \over \tilde{\zeta}_i}$ ):
$$\eqalign{
\prod_{i=0}^{1} & (- \tilde{s}_2 +2 \tilde{s}_3 -i) = 0 \cr
\prod_{i=1}^{3} & (- \tilde{s}_1 +i/3) =0 \cr
\prod_{i=0}^{1} & (-\tilde{s}_1 +2 \tilde{s}_2 -i) =0~~. \cr}
\eqlabel{indeqsol}$$
The above system has $12$ solutions. It was obtained by acting with
the system of operators~\eqref{d3} on
$\tilde{F}(\tilde{\zeta}_i)$ given in the following form:
$$
 \tilde{F}(\tilde{\zeta}_i)~ =~
\tilde{\zeta}_1^{\tilde{s}_{1}}
\tilde{\zeta}_2^{\tilde{s}_{2}}
\tilde{\zeta}_3^{\tilde{s}_{3}}
\sum_{a,b,c=0}^{\infty}C_{abc}
\tilde{\zeta}_1^a\tilde{\zeta}_2^b
\tilde{\zeta}_3^c \quad \hbox{ for }~
\tilde{\zeta}_i \rightarrow 0 ~~.
\eqlabel{indeq}
$$

The other method of counting the number of solutions is by calculating
the volume of the dual cone \cite{\rgelfand}.
When the dual polyhedron is a simplex, the volume is the absolute value
of the determinant of the generators of the dual cone (the volume of
the $n$-dimensional cube is taken to be $n!$).
Otherwise, we can use the simplicial decomposition used to find the
generators of the Mori cone, and then add up the volumes of the
different simplices.

Taking the last two differential operators from \eqref{d3}, but dropping the
$\Theta_3$ term in the
second, we are left with the $\Delta^*$ hypergeometric system for
the $K3$ hypersurface
\cp{(1,1,2,2)}[6].
To see this we have to do the same type of analysis for
\cp{(1,1,2,2)}[6].
The dual polyhedron contains $6$ points:
$$\eqalign{
 \bar{\nu}_0^* &= (\- 1,\- 0,\- 0,\- 0) \cr
 \bar{\nu}_1^* &= (\- 1, - 1, - 1, - 1) \cr
 \bar{\nu}_2^* &= (\- 1,\- 1, - 1, - 1) \cr
 \bar{\nu}_3^* &= (\- 1,\- 0,\- 1,\- 0) \cr
 \bar{\nu}_4^* &= (\- 1,\- 0,\- 0,\- 1) \cr
 \bar{\nu}_5^* &= (\- 1,\- 0, - 1, - 1) \cr}
$$
out of which one  is interior
to a $1$-face, so we are going to study the family:
$$
 p = \sum_{i=1}^{4} a_i x_i^{6 \over k_i} + a_0 \prod_{i=1}^{4} x_i +
a_5 x_1^3 x_2^3
$$
The Mori cone has two generators:
$$\eqalign{
{\bf l}^1 &= ( - 3,\- 0,\- 0,\- 1,\- 1,\- 1)\cr
{\bf l}^2 &= (\- 0,\- 1,\- 1,\- 0,\- 0, -2)\cr}
$$
The large complex structure parameters are:
$$\eqalign{
\zeta_1 &= -{a_3 a_4 a_5 \over a_0^3} \cr
\zeta_2 &= {a_1 a_2 \over a_5^2} \cr}
\eqlabel{zeta2}
$$
In terms of $\theta_1 = \zeta_1 {\partial\over\partial \zeta_1}$ and
$\theta_2 = \zeta_2 {\partial\over\partial \zeta_2}$, the
differential operators are:
$$\eqalign{
{\cal D}_1 &= \theta_1^2 (\theta_1 -2 \theta_2)- 3^3 \zeta_1 \prod_{i=1}^3
(\theta_1 + i/3) \cr
{\cal D}_2 &= \theta_2^2 - \zeta_2 \prod_{i=0}^1 (\theta_1 -2 \theta_2
-i)\cr}
\eqlabel{d2}
$$
The system \eqref{d2} can be shown to have 6 linearly independent solutions.
We see that in the limit $\zeta_3 \rightarrow 0$ the system
\eqref{d3} reduces to \eqref{d2}, while if we also let $\zeta_2
\rightarrow  0$ the operator ${\cal D}_1$ reduces to the $\cal D$ operator in
\eqref{Dtorus}.

Obviously, one would expect the solutions for these hypergeometric systems to
obey the same rule, \ie taking the limit $\zeta_2\rightarrow 0$ in the
semi-period expression for $K3$, one would get the semi-period expresion for
$T^2$. Similarly, the limit $\zeta_3\rightarrow 0$ in the semi-periods for
\cp{(1,1,2,4,4)}[12] will yield $K3$ semi-periods. To show this, we
will analytically continue our expressions to the regions where the desired
parameter is small, and then take the zero limit of that parameter.

First, we need to calculate semi-periods.
Working in the region of the moduli space where $p_0$ represents the Fermat
part of the defining polynomial,
and dropping some numerical factors, one has the following:\hfil\break
--for the torus
$$ \eqalign{
  F_{T^2}
           &= {1\over a_0}\, \sum_{m=0}^\infty\,
         \l^{(\d_2+\d_3)(m+1)}\,
         \left({1 \over \zeta_1}\right)^{{m+1\over 3}}
         {(-1)^m \over m!}\,
         \Gamma^3 \left({m+1\over 3}\right)~;\cr}
\eqlabel{spt2}$$
--for the $K3$
$$ \eqalign{
  F_{K3}
           &= {1\over a_0}\, \sum_{m=0}^\infty
         \l^{(\d_2+2\d_3+2\d_4)(m+1)}\,
         \left({1 \over \zeta_1}\right)^{{m+1\over 3}}
         {(-1)^m\over m!}\, \Gamma^2 \left({m+1\over 3}\right)\cr
&\hskip10pt \times
         \left({1 \over \zeta_2}\right)^{{m+1\over 6}}
        \sum_{n=0}^\infty
         \l^{3 n \d_2}\,
         \left({1 \over \zeta_2}\right)^{{n\over 2}}
         {(-1)^n\over n!}
       \Gamma^2 \left({m+3n+1\over 6}\right)~;\cr}
\eqlabel{spk3}$$
--for the Calabi--Yau
$$\eqalign{
  F_{CY}
           &= {1\over a_0}\, \sum_{m=0}^\infty
         \l^{(\d_2+2\d_3+4\d_4+4\d_5)(m+1)}\,
         \left({1 \over \zeta_1}\right)^{{m+1\over 3}}
         {(-1)^m\over m!}\, \Gamma^2 \left({m+1\over 3}\right)\cr
&\hskip10pt \times
         \left({1 \over \zeta_2}\right)^{{m+1\over 6}}
      \sum_{n=0}^\infty
         \l^{3 n (\d_2+2\d_3)}\,
         \left({1 \over \zeta_2}\right)^{{n\over 2}}
         {(-1)^n\over n!}
       \Gamma \left({m+3n+1\over 6}\right)\cr
&\hskip10pt \times
        \left({1 \over \zeta_3}\right)^{{m+3n+1\over 12}}
      \sum_{p=0}^\infty
         \l^{6 p \d_2}\,
        \left({1 \over \zeta_3}\right)^{{p\over 2}}
         {(-1)^p\over p!}
             \Gamma^2 \left({m+3n+6p+1\over 12}\right)~.\cr}
\eqlabel{spcy}$$
Here $\l = e^{2\pi i \over d}$,
$d$ being the degree of the polynomial (3, 6 or 12 respectively),
and $\d_i = 0,1,...{d \over k_i}-1$.
This means our contours connect the
points $0,{\l^{k_i\d_i}} \infty$ in the $x_i$ plane.
Not all different chains lead to different semi-periods;
{}from the above equations one can see the number of linearly independent
semi-periods that can be obtained is 3 for the
torus, 6 for $K3$ and 12 for the three-dimensional \cym .
These then are all the solutions of the respective hypergeometric equations.
To see that the solutions \eqref{spcy} are of the type
\eqref{indeq}, one can break up the summations in \eqref{spcy}
over $m,$ $n$ and $p$
such that
 $m = 3\tilde m + i$ for $i=0,1,2,$ $n = 2\tilde n + j$ for
$j=0,1,$ and $p = 2\tilde p + k$ for $k=0,1$.
Now the summation over $m$
becomes a summation from $0$ to $\infty$ over $\tilde m$
and a summation from $0$ to $2$ for $i$,
and likewise for the others.
Now it is clear that the semi-period separates into
12 summations with different leading order behaviours as
the $1/\zeta_i \rightarrow 0$.
These leading orders are those predicted by the
solutions of \eqref{indeqsol}.

{}For convenience in doing the analytic continuation
we set $\d_2=0$ and focus on the $\zeta_3$ dependent part of \eqref{spcy}.
Its Barnes integral representation is as follows
{}~\Ref{\rSGI}{G.~Sansone and J.~Gerretsen, {\it Lectures on the Theory of
Functions of a Complex Variable} (Wolters-Noorhoff, Groningen 1969, the
Netherlands).}:
$$ \eqalign{
   {-1\over 2 \pi i} \int_0^{i \infty}
\rd s \left({1\over \zeta_3}\right)^{{m+3n+6s+1\over 12}}\Gamma(-s)
\Gamma^2 \left({m+3n+6s+1\over 12}\right)~.\cr}
\eqlabel{Barnes}$$
The poles of $\Gamma(-s)$ lie along the positive axis. If the contour of
integration encloses these poles, then we obtain our starting expression,
since for $s\rightarrow -p$
$${\rm Res}\, \Gamma(s) ={(-1)^{p+1}\over p!}~. $$
The poles of $\Gamma^2 \left({m+3n+6s+1\over 12}\right)$ lie at
${m+3n+6s+1\over 12} = -k$, where $k = 0,1,2,3...$. If we enclose these poles
in the integrating contour (\ie close the contour to the left)
the value of the integral \eqref{Barnes} will be:
$${- \sum_{k=0}^\infty \Gamma\left({12k+m+3n+1\over 6}\right) \left({1 \over
\zeta_3}\right)^{-k} \left({1 \over k!}\right)^{2} 2 \ps_{1}(1+k)~.}
\eqlabel{ksum}$$
Here \cite{\rErdelyi} \
$$ \ps_{1}(k+1)=1+{1\over 2}+{1\over 3}+...+{1\over k}-\g~,$$
and $\g$ is the Euler-Mascheroni constant.
We have also used the fact that
$$\Gamma(-k+\e)={(-1)^k \over k!}{({1\over \e}+{\ps_{1}(1+k)}+{\it O} (\e))}$$
Thus we see that the limit $\zeta_3 \rightarrow 0$ in \eqref{spcy} will give
an expression proportional to \eqref{spk3}.

In a similar manner, one can show that
by analytically continuing \eqref{spk3} to small $\z_2$,
the limit $\zeta_2 \rightarrow 0$  is proportional to \eqref{spt2}.

 %
\vfill\noindent
{\bf Acknowledgments}:
The authors gratefully acknowledge much input and
guidance from Philip Candelas. We also want to thank Tristan H\"ubsch
and Mihaela Manoliu for interesting discussions.
A.C.A. and D.~J. were supported by
NSF grant PHY 9009850, PHY 9511632 and the Robert~A.~Welch Foundation.
E.~D. was supported by the Alexander von Humboldt Foundation.
E.~D. would like to thank  Howard University for hospitality while part of this
work was germinating.
Some of this work was carried out while E.~D. was a member of
the Theory Group at the University of Texas.

\vfill\eject

\immediate\closeout\referencewrite
\referenceopenfalse
\line{\bf\hfil References\hfil}\vskip.2truein
\input referenc.texauxil

\bye